\documentstyle[aps,prd,epsfig,floats,axodraw]{revtex}

\newcommand{\beq}{\begin{equation}}
\newcommand{\eeq}{\end{equation}}
\newcommand{\bea}{\begin{eqnarray}}
\newcommand{\beas}{\begin{eqnarray*}}
\newcommand{\eea}{\end{eqnarray}}
\newcommand{\eeas}{\end{eqnarray*}}
\newcommand{\ba}{\begin{array}}
\newcommand{\ea}{\end{array}}

\def\lsim{\:\raisebox{-0.75ex}{$\stackrel{\textstyle<}{\sim}$}\:}

\begin{document}

\title{Heavy particle production during reheating}

\author{Rouzbeh Allahverdi~$^{1,2}$ and Manuel Drees~$^{1}$}

\address{$^{1}$ Physik Department, TU Muenchen, James Frank
Strasse, D-85748, Garching, Germany. \\
$^{2}$ Theory Group, TRIUMF, 4004 Wesbrook Mall, Vancouver, B.C., V6T 2A3, 
Canada,.}

\maketitle

\begin{abstract}
We discuss production of heavy partciles during reheating. We find that
the very energetic inflaton decay products can contribute to the
production of massive stable particles, either through collisions with
the thermal plasma, or through collisions with each other. If such
reactions exist, the same massive particles can also be produced
directly in inflaton decay, once higher--order processes are
included. We show that these new, non--thermal production mechanisms
often significantly strengthen constraints on the parameters of models
containing massive stable particles.
\end{abstract}

\vskip2pc
\section{Introduction}

After inflation \cite{infl}, coherent oscillations of the inflaton dominate 
the energy density of the Universe.
At some later time these coherent oscillations decay to the fields to
which they are coupled, and their energy density is transferred to
relativistic particles; this reheating stage results in a
radiation--dominated Friedmann--Robertson--Walker (FRW) universe.

Until a few years ago, reheating was treated as the perturbative, one
particle decay of the inflaton with decay rate $\Gamma_{\rm d}$, leading to 
the simple estimate
$T_{\rm R} \sim {({\Gamma}_{\rm d}{M}_{\rm Planck})}^{1/2}$ for the
reheat temperature \cite{reheat}, where $M_{\rm Planck}
= 2.4 \times 10^{18}$ GeV represents the reduced Planck mass. It has been 
noticed in recent years that the initial stage of inflaton
decay might occur through a complicated and non--perturbative process
called parametric resonance \cite{preheat2}. However, it is
generally believed that an epoch of (perturbative)
reheating from the decay of massive particles (or coherent field
oscillations, which amounts to the same thing) is an essential
ingredient of any potentially realistic cosmological model
\cite{jed}. In what follows we generically call the decaying particle
the ``inflaton''. However, it should be clear that our results are equally
well applicable to any other particle whose (late) decay results in
entropy production.

Even before all inflatons decay, the decay products form a plasma
which has the instantaneous temperature 
\beq \label{temp}
T \sim \left( g_*^{-1/2} H \Gamma_{\rm d} M^{2}_{\rm Planck}
\right)^{1/4}, 
\eeq
where $H$ is the Hubble parameter and $g_*$ denotes the number of
relativistic degrees of freedom in the plasma \cite{kt}. This temperature 
reaches its 
maximum $T_{\rm max}$ soon after the inflaton field starts to oscillate. In 
addition to this thermalized plasma there are inflaton decay products with 
energy $\simeq m_\phi/2$, which will eventually come into equilibrium with 
the thermal bath.

In recent years several mechanisms have been put forward for creating very
heavy, even superheavy, particles in cosmologically interesting
abundances \cite{ckr1}. Here we will focus on production of very
massive particles from various processes, including a thermal bath,
during perturbative reheating. Note that particle production from
other sources, if present, would further strengthen the bounds which
we will derive as they simply add to production from mechanisms
discussed here.

\section{Heavy particle production}

In Ref. \cite{ckr2} out of equilibrium production of $\chi$ from
scatterings of ``soft'' particles in the thermal bath (with energy $E \sim 
T$) is studied and the final result is
found to be (the superscript ``ss'' stands for $\chi$ production from
``soft--soft'' scattering)
\bea  \label{ssrate1}
\Omega^{\rm ss}_\chi h^2 &\sim& \left ({200 \over g_*} \right )^{3/2}
\alpha_\chi^2 {\left ({2000 T_{\rm R} \over m_\chi} \right )}^7 
\nonumber \\ && \hspace*{2cm}
(\chi \ {\rm not \ in \ equilibrium}).
\eea
In the opposite situation, with $\chi$ being initially in equilibrium with the 
thermal bath, today's $\chi$ relic density is given \cite{gkr}
by
\bea \label{ssrate2}
\Omega^{\rm ss}_\chi h^2 &\sim& \left( \frac {200} {g_*} \right)^{1/2}
\frac {T_{\rm R} x_{\rm f}^{4+a}} { m_\chi \alpha_\chi^2}  \left( \frac
{T_{\rm R}} { 8 \cdot 10^5 \ {\rm GeV}} \right)^2 \nonumber \\ &&
\hspace*{2cm} (\chi \ {\rm in \ equilibrium}) ,
\eea
where the exponent $a = 0 \ (1)$ if $\chi \chi$ annihilation proceeds
from an $S$ ($P$) wave initial state. The freeze--out temperature is
now given by $x_{\rm f} \equiv m_\chi/ T_{\rm f} \simeq \log ( 0.08
g_*^{-1/2} \alpha_\chi^2 x_{\rm f}^{2.5-a} M_{\rm Planck} T^2_{\rm R}
/ m^3_\chi)$. 

However, as mentioned earlier, ``hard'' particles with energy $E \simeq 
m_\phi/2 \gg T$ are continuously
created by inflaton decay for $H \geq \Gamma_{\rm
d}$. These particles eventually thermalize with the bath, but this
takes a finite amount of time. The presence of hard inflaton decay
products can therefore affect heavy particle production in two
ways. Firstly, $\chi$s can be produced from $2 \rightarrow 2$
scatterings of a hard particle off either soft particles in the
thermal bath (if kinematically allowed), or off other hard 
particles \cite{ad1}. Moreover, $\chi$s might be directly produced from 
inflaton decay \cite{ad2}.

\subsection{Particle production from hard--soft scatterings}

In order to estimate the rate of heavy particle production from the
``hard'' inflaton decay products, we also have to know the time needed
to reduce their energy from a value $\sim
m_\phi/2$ to a value near $T$. As shown in Ref. \cite{ds}, $2
\rightarrow 2$ scattering reactions are {\em not} very efficient in this 
respect. The
reaction rate is large, but the average energy loss per scattering is
only ${\cal O}(T^2/m_\phi)$, giving a slow--down time of order $
\left[ \alpha^2 T^2 / m_\phi \right]^{-1}$ (up to logarithmic
factors). On the other hand,
inelastic $2 \rightarrow 3$ reactions allow large energy losses (in
nearly collinear particles) even if all virtual particles only have
virtuality of order $T$. The slow--down rate is thus given by:
\beq \label{slowrate}
\Gamma_{\rm slow} \simeq 3 \alpha^3 T \left( \frac {g_*} {200}
\right)^{1/3}. 
\eeq

Next let us estimate the rate for $\chi$ pair production from
hard--soft scatterings. This process is kinematically allowed so long
as $ET \geq 4 m^{2}_{\chi}$, where $E$ is the energy of the hard
particle so that the square of the center--of--mass energy is
typically a few times $ET$. The hard particle initially has energy $E
\simeq m_\phi/2$ and average number density $\bar n_h \sim
g_* T^4/(3 m_\phi)$, just after its production from inflaton decay.
On the other hand, the rate for $\chi$ production from hard--soft
scatterings is approximately given by
\beq \label{hsprod}
\Gamma^{\rm hs}_\chi \sim \left( \frac {\alpha^2_\chi} {T m_\phi}
+ \frac {\alpha \alpha_\chi^2} {m_\chi^2} \right) 0.2 T^3.
\eeq 
The two
contributions in (\ref{hsprod}) describe $2 \rightarrow 2$ reactions
with squared center--of--mass energy $\sim m_\phi T$ and ``radiative
return'' $2 \rightarrow 3$ reactions, respectively; in the latter
case the hard particle emits a collinear particle prior to the
collision, thereby reducing the effective cms energy of the collision
to a value near $m_\chi$. In order to make a safe (under)estimate we
choose the temperature $T_0 = 2 T_{\rm thr}$ for presenting our
results; note that the $\chi$ pair production cross section at
threshold, $s = 4 m^2_\chi$, is suppressed kinematically.

Our final
results for the contribution of hard--soft collisions to the $\chi$
relic density are \cite{ad1}: for $T_0 < T_{\rm R}$:
\bea \label{hsrate3}
\Omega_\chi^{\rm hs} h^2 \sim && \left( \frac {200}{g_*}
\right)^{1/3} \frac {\alpha_\chi^2}{\alpha^2} \left( \frac {T_{\rm R}}
{10^4 \ {\rm GeV} } \right)^2 \frac {10^{13} \ {\rm GeV}} {m_\phi}
\nonumber \\ &\cdot&
\frac {100 T_{\rm R}} {m_\chi} \left( 1 + \frac {m^2_\chi} {\alpha
T_{\rm R} m_\phi} \right), \ \ \ (T_0 < T_{\rm R})
\eea
in the opposite situation, we have
\bea \label{hsrate4}
\Omega_\chi^{\rm hs} h^2 &\sim& \left( \frac {200}{g_*} \right)^{1/3}
\frac {\alpha^2_\chi} {\alpha^3} \frac {m_\phi} {10^{13} \ {\rm GeV}}
\left( \frac {3000 T_R} {m_\chi} \right)^5, \nonumber \\
&& \hspace*{2cm} (T_0 > T_{\rm R}) 
\eea
%

\subsection{Particle production from hard--hard scatterings}

If $T_0 > T_{\rm R}$, we should also consider $\chi$ production from
scattering of two hard particles. Collisions of
these particles with each other can produce $\chi$ pairs if $m_\chi <
m_\phi / 2$. Note that this constraint is independent of the
temperature. On the
other hand, it's also possible that $T_0 > T_{\rm max}$, in which case
hard--soft scattering (and soft--soft scattering) does not produce any
$\chi$ particles.

The rate of $\chi$ production from hard--hard scattering is
quadratic in the density of hard particles. Therefore we can not
use our earlier approximation of the
density $\bar n_h$ of hard particles produced in one Hubble time in the 
presence of a thermalized plasma, since the 
actual density $n_h(t)$ at any given time
will be much smaller than this. In a plasma with temperature $T$ a hard 
particle will only survive for a time 
$\sim 1/\Gamma_{\rm slow}$, see eq.(\ref{slowrate}). For $T_{\rm max} > T > 
T_{\rm R}$, the production of hard
particles from inflaton decays and their slow--down will be in
equilibrium, i.e. the instantaneous density $n_h(t) = 2 \Gamma_d
n_\phi(t) / \Gamma_{\rm slow}$, where $n_\phi$ is the density of
inflatons. By taking into account $\chi$ production prior to the bulid up of a 
thermalized plasma, our final estimate will be \cite{ad1}
\beq \label{hhrate}
\Omega_\chi^{\rm hh} h^2 \sim 6 \cdot 10^{27} \cdot \left( \frac
{g_*}{200} \right)^{1/2} \sigma_{\chi \chi} \frac {m_\chi T_{\rm R}^7}
{m^2_\phi T^4_{\rm max}}.
\eeq

In Fig. (2) we present three numerical examples to compare the significance 
of hard-soft and hard-hard scatterings with that of soft-soft scatterings. A 
detailed discussion on this figure and features observed in it can be found 
in Ref. \cite{ad1}.

\subsection{Particle production from inflaton decay}

We now discuss the direct production of $\chi$ particles in inflaton
decay whose importance has recently been noticed \cite{ad2}.  Let us denote 
the average number of $\chi$ particles
produced in each $\phi$ decay by $B(\phi \rightarrow \chi)$. The
$\chi$ density from $\phi$ decay can then be estimated as: 
\beq \label{phidec}
\Omega^{\rm decay}_\chi h^2 \simeq 2 \cdot 10^8 B(\phi \rightarrow
\chi) \frac {m_\chi} {m_\phi} \frac {T_{\rm R}} {1 \ {\rm GeV}}.
\eeq
Eq. (\ref{phidec}) holds if the $\chi$ annihilation rate is smaller
than the Hubble expansion rate at $T \simeq T_{\rm R}$. 

We now discuss 
estimates of $B(\phi \rightarrow \chi)$. This
quantity is obviously model dependent, so we have to investigate
several scenarios. The first, important special case is where $\chi$
is the LSP. If $m_\phi$ is large compared to typical visible--sector
superparticle masses, $\phi$ will decay into particles and
superparticles with approximately equal probability \cite{ad2}. Moreover, all 
superparticles will quickly decay into the LSP and some standard particle(s). 
As a result, if $\chi$ is
the LSP, then $B(\phi \rightarrow \chi) \simeq 1$, independently of
the nature of the LSP.

Another possibility is that the inflaton couples to all particles with
more or less equal strength, e.g. through non--renormalizable
interactions. In that case one expects $B(\phi \rightarrow \chi) \sim
1/g_* \sim 1/200$. However, even if $\phi$ has no direct couplings to
$\chi$, the rate (\ref{phidec}) can be large. The key observation is
that $\chi$ can be produced in $\phi$ decays that occur in higher
order in perturbation theory whenever $\chi$ can be produced from
annihilation of particles in the thermal plasma. In most realistic
cases, $\phi \rightarrow f \bar f \chi \bar \chi$ decays will be
possible
\vspace*{6mm}
\begin{center}
\SetScale{0.6} \SetOffset(40,40)
\begin{picture}(225,125)(0,0)
\DashLine(0,50)(75,50){5} \Text(0,25)[l]{$\phi$}
\Vertex(75,50){3}
\ArrowLine(125,25)(75,50) \Text(80,16)[h]{$\bar{f}$}
\ArrowLine(75,50)(125,75) \Text(60,50)[t]{$f$}
\Vertex(125,75){3}
\Photon(125,75)(175,100){5}{4} 
\Vertex(175,100){3}
\ArrowLine(125,75)(175,50) \Text(110,30)[h]{$f$}
\ArrowLine(175,100)(225,75) \Text(145,45)[r]{$\chi$}
\ArrowLine(225,125)(175,100) \Text(145,75)[r]{$\bar \chi$}              
\end{picture}
\vspace*{-13mm}

\noindent
{\bf FIG. 1:}~Sample diagram for $\chi$ production in four-body
inflaton decay. 
\end{center}
\vspace*{3mm}
if $\chi$ has gauge interactions, where $f$ stands for some
gauge non--singlet with tree--level coupling to $\phi$. A diagram
contributing to this decay is shown in Fig.~(1). Note that the part of
the diagram describing $\chi \bar \chi$ production is identical to the
diagram describing $\chi \bar \chi \leftrightarrow f \bar f$
transitions. This leads to the following estimate:
\beq \label{fourbody}
B(\phi \rightarrow \chi)_4 \sim \frac {C_4 \alpha_\chi^2} {96 \pi^3}
\left( 1 - \frac {4 m_\chi^2} {m_\phi^2} \right)^2 \left( 1 - \frac {2
m_\chi} {m_\phi} \right)^{5 \over 2},
\eeq
where $C_4$ is a multiplicity (color) factor. The phase space factors
have been written in a fashion that reproduces the correct behavior
for $m_\chi \rightarrow m_\phi/2$ as well as for $m_\chi \rightarrow
0$. This estimate provides a lower bound on $B(\phi \rightarrow \chi)$
under the conditions assumed for our calculation of $\Omega_\chi^{\rm
hs}$ and $\Omega_\chi^{\rm hh}$; whenever a primary inflaton decay
product can interact with a particle in the thermal plasma, or with
another primary decay product, to produce a $\chi \bar \chi$ pair, $\phi
\rightarrow \chi$ four--body decays must exist.

Occasionally one has to go to even higher order in perturbation
theory to produce $\chi$ particles from $\phi$ decays. For example, if
$\chi$ has only strong interactions but $\phi$ only couples to $SU(3)$
singlets, $\chi \bar \chi$ pairs can only be produced in six body
final states, $\phi \rightarrow f \bar f q \bar q \chi \bar \chi$. A
representative diagram can be obtained from the one shown in Fig.~(1) by
replacing the $\chi$ lines by quark lines, attaching an additional
virtual gluon to one of the quarks which finally splits into $\chi
\bar \chi$. The branching ratio for such six body decays can be
estimated as
\beq \label{sixbody}
B(\phi \rightarrow \chi)_6 \sim \frac {C_6 \alpha_\chi^2 \alpha^2}
{1.1 \cdot 10^7}
\left( 1 - \frac {4 m_\chi^2} {m_\phi^2} \right)^4 \left( 1 - \frac {2
m_\chi} {m_\phi} \right)^{9 \over 2}.
\eeq
Finally, in supergravity models with explicit (supersymmetric) $\chi$
mass term there in general exists a coupling between $\phi$ and either
$\chi$ itself or, for fermionic $\chi$, to its scalar superpartner, resulting 
in the estimate \cite{aem}
\beq \label{gravbr}
B(\phi \rightarrow \chi) \sim \frac {v^2 m^2_\chi m_\phi} {16 \pi
\sqrt{g_*} M^{3}_{\rm Planck} T^2_{\rm R} } \left( 1 - \frac {4 m_\chi^2}
{m_\phi^2} \right)^{1 \over 2},
\eeq
where $v$ denotes the vacuum expectation value of the inflaton at the true 
minimum of its potential.

\section{Discussion}

The production of $\chi$ particles from inflaton decay will be
important for large $m_\chi$ and large ratio $m_\chi / T_{\rm R}$, but
tends to become less relevant for large ratio $m_\phi / m_\chi$. Even
if $m_\chi < T_{\rm max}$, $\chi$ production from the thermal plasma
(\ref{ssrate1}) will be subdominant if
\beq \label{compare1}
\frac {B(\phi \rightarrow \chi)} {\alpha_\chi^2} > \left( \frac {100
T_{\rm R}} {m_\chi} \right)^6 \frac {m_\phi} {m_\chi} \frac {1 \ {\rm
TeV}} {m_\chi}.
\eeq
Note that the first factor on the r.h.s. of (\ref{compare1}) must be $\lsim
10^{-6}$ in order to avoid over--production of $\chi$ from thermal
sources alone.

In \cite{ad2} we showed that the decay contribution (\ref{phidec}) by
itself leads to very stringent constraints on models with massive
stable $\chi$ particles. In particular, charged stable particles with
mass below $\sim 100$ TeV seem to be excluded, unless $m_\chi >
m_\phi/2$. In case of a (neutral) LSP with mass around 200 GeV, the
overclosure constraint implies $m_\phi / T_{\rm R} > 4 \cdot 10^{10}$,
i.e. a very low reheat temperature, unless $\chi$ was in thermal
equilibrium below $T_R$; recall that $B(\phi \rightarrow \chi) = 1$ in
this case. Finally, if $m_\phi \sim 10^{13}$ GeV a ``wimpzilla'' with
mass $m_\chi \sim 10^{12}$ GeV will be a good Dark Matter candidate
only if it has a very low branching ratio, $B(\phi \rightarrow \chi)
\sim 5 \cdot 10^{-8} \ {\rm GeV} / T_{\rm R}$, i.e. if its couplings
to ordinary matter are very small.

Many of the results presented here are only
semi--quantitative. Unfortunately in most cases significant
improvements can only be made at great effort. For example, a proper 
treatment of the slow--down of
primary inflaton decay products would require a careful treatment of
the full momentum dependence of the particle distribution
functions. On the other hand, our estimates of $B(\phi \rightarrow \chi)$
should be quite reliable if $m_\chi > T_{\rm R}$
(which is required for $\chi$ not to have been in thermal equilibrium
at $T_{\rm R}$); even for many--body decays, details of the matrix
elements should change our estimates only be ${\cal O}(1)$
factors. Fortunately this is often also the most
important of the new mechanisms for the production of
massive particles at the end of inflation.

\section*{Acknowledgements}
This work was supported by ``Sonderforschungsbereich 375
f\"ur Astro-Teilchenphysik'' der Deutschen Forschungsgemeinschaft.


\newpage
\setcounter{figure}{1}
\begin{figure}
\hspace*{3cm}
\vspace*{-0.7cm}
\epsfig{figure=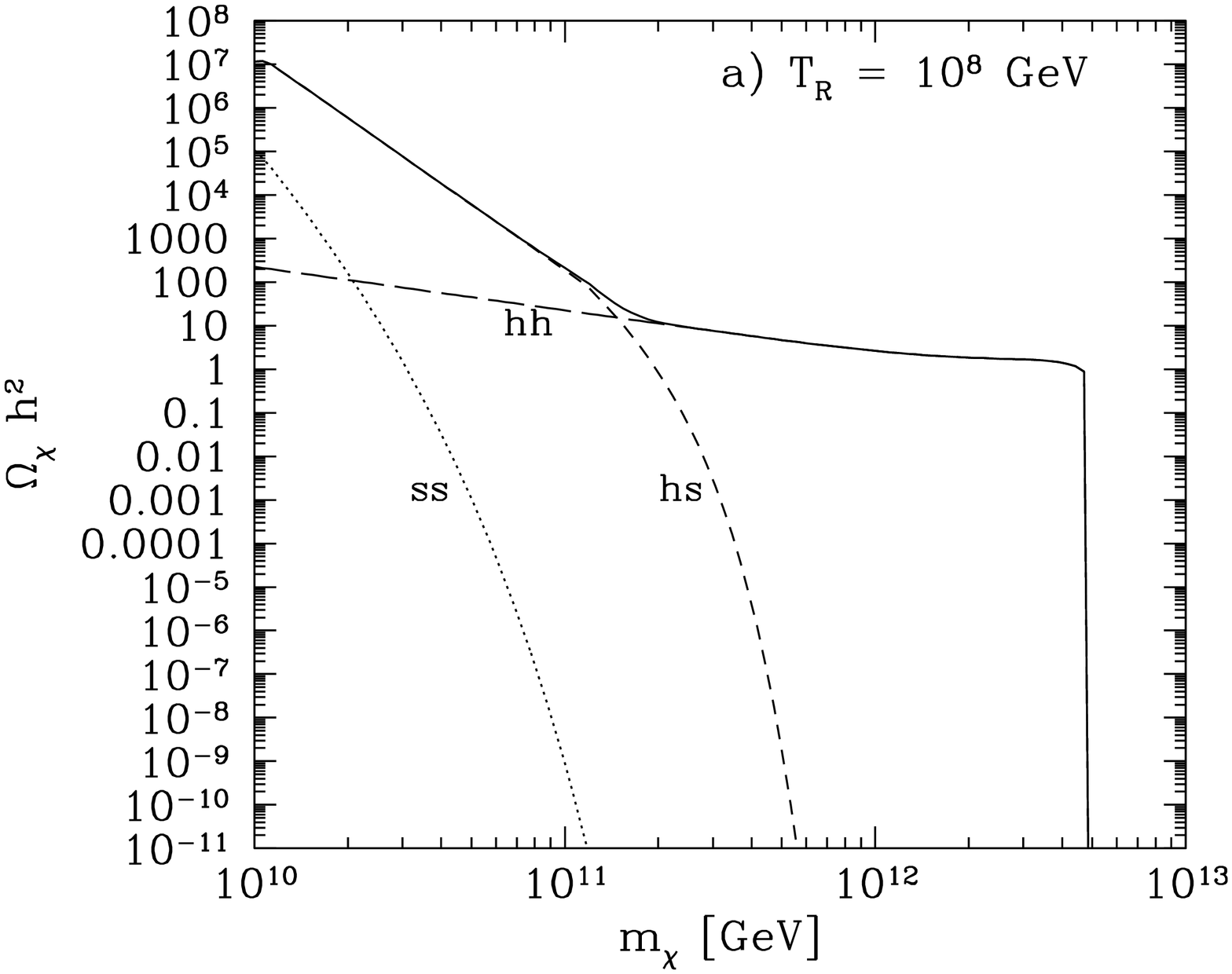,angle=-90,width=0.56\textwidth,clip=}

\hspace*{3cm}
\vspace*{-.7cm} 
\noindent
\epsfig{figure=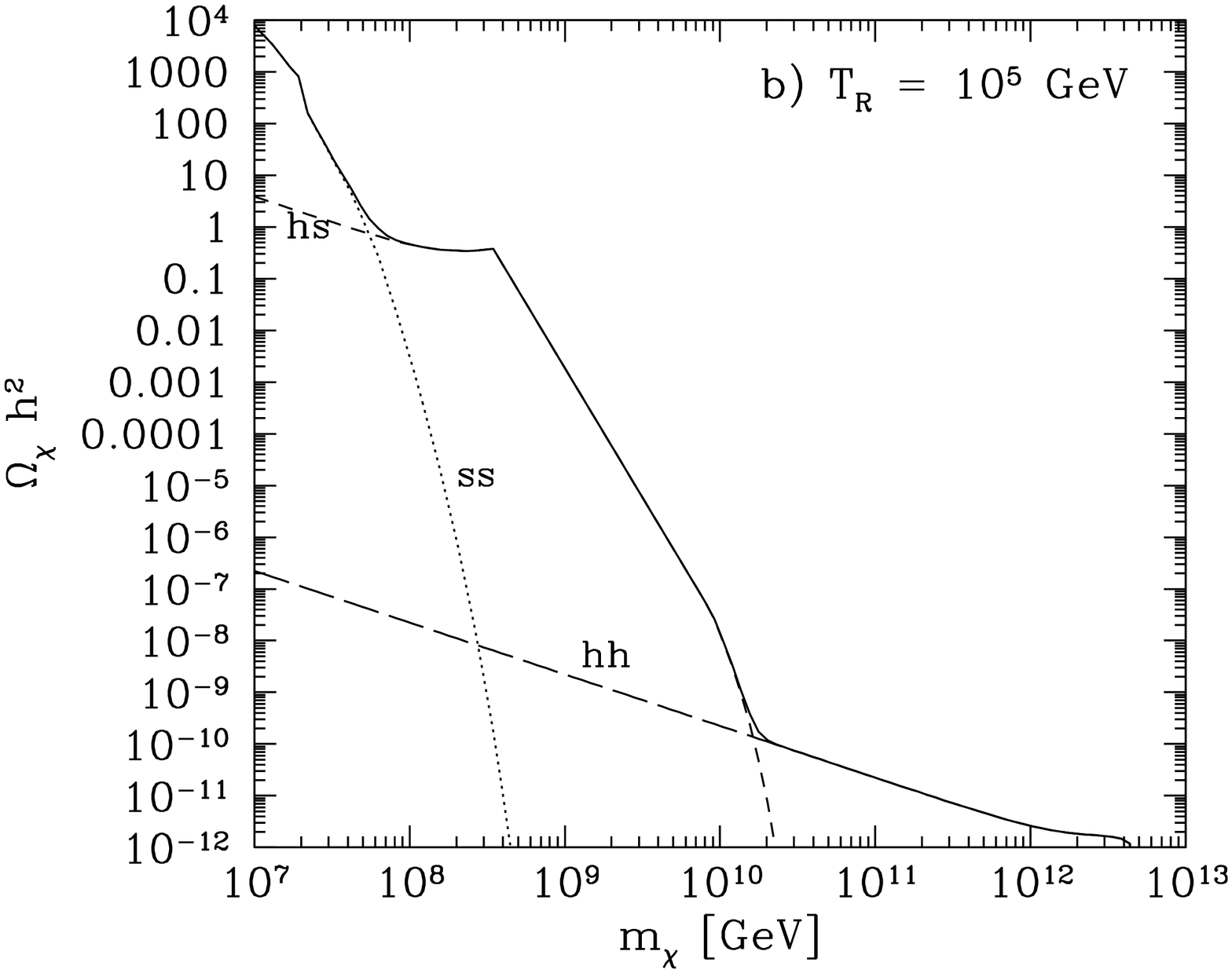,angle=-90,width=0.56\textwidth,clip=}

\hspace*{3cm}
\vspace*{-.1cm} 
\noindent
\epsfig{figure=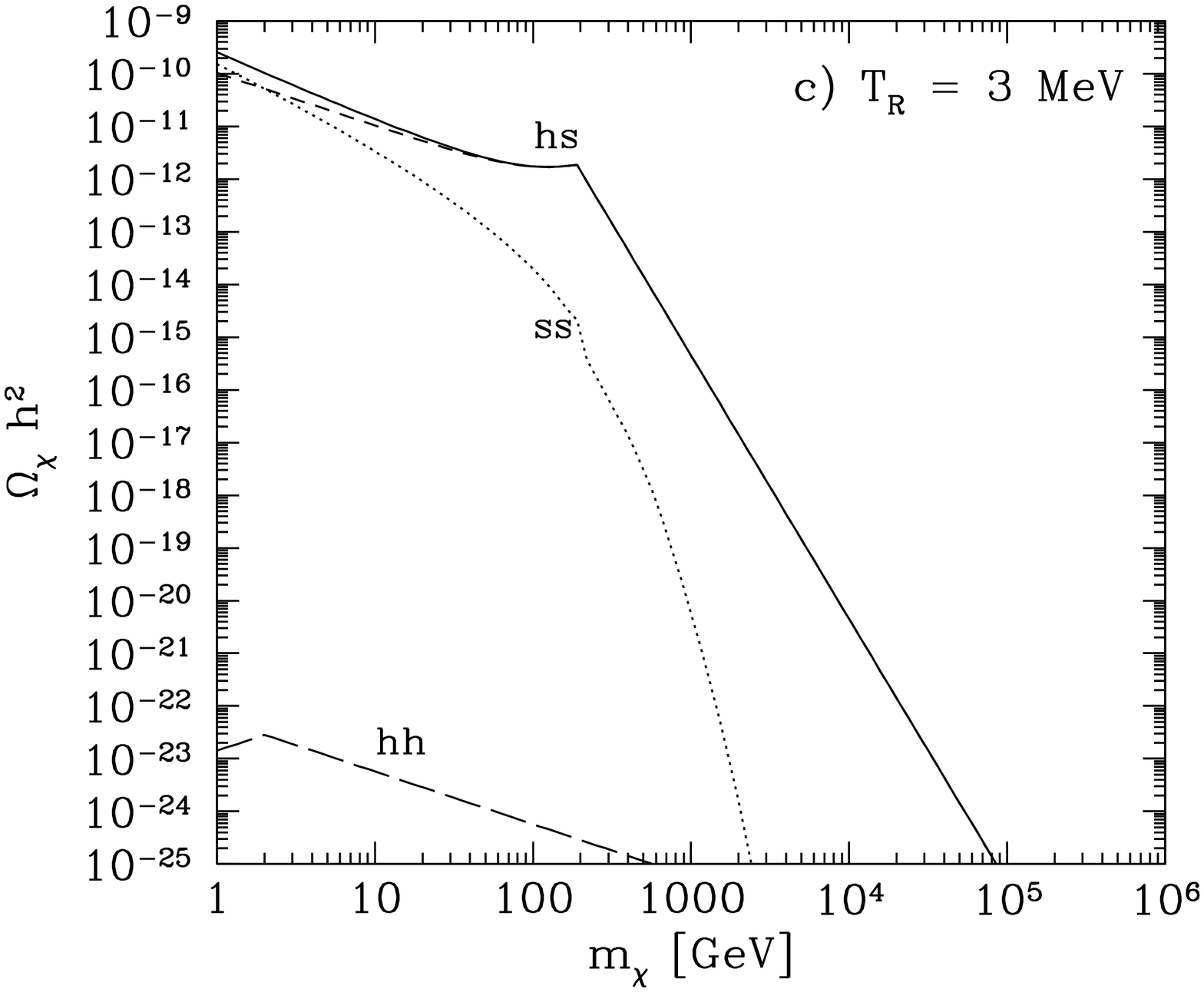,angle=-90,width=0.56\textwidth,clip=}

\caption{The relic density $\chi$ particles would currently have if
they are absolutely stable is shown as a function of $m_\chi$
for couplings $\alpha = 0.05, \ \alpha_\chi = 0.01$, and a) $(T_{\rm R},
m_\phi) = (10^8 \ {\rm GeV}, 10^{13} \ {\rm GeV})$, b)$(10^5 \ {\rm
GeV}, 10^{13} \ {\rm GeV})$, c)$ (3 \ {\rm MeV}, 10^{8} \ {\rm
GeV})$. The soft--soft, hard--soft and hard--hard contributions are
shown by the dotted, short dashed and long dashed curves,
respectively, while the solid curves show the sum of all three
contributions.} 
\end{figure}

\end{document}